\documentclass[11pt,twoside]{article}
\usepackage{./asp2014}
\newcommand{\msun}{\mbox{M$_{\odot}$}}
\newcommand{\microjy}{\mbox{$\mu$Jy}}

\aspSuppressVolSlug
\resetcounters

\begin{document}

\title{Exploring Protostellar Disk Formation with the ngVLA}
\author{John J. Tobin}
\affil{University of Oklahoma, \email{jjtobin@ou.edu}}
\author{Patrick Sheehan}
\affil{University of Oklahoma}
\author{Doug Johnstone}
\affil{ NRC-Herzberg}
\author{Rajeeb Sharma}
\affil{University of Oklahoma}

\paperauthor{John J. Tobin}{jjtobin@ou.edu}{}{University of Oklahoma}{}{Norman}{OK}{73072}{USA}
\paperauthor{Patrick D. Sheehan}{sheehan@ou.edu}{}{University of Oklahoma}{}{Norman}{OK}{73072}{USA}
\paperauthor{Doug Johnstone}{doug.johnstone@gmail.com}{}{NRC-Herzberg}{}{Victoria}{BC}{}{CA}
\paperauthor{Rajeeb Sharma}{Rajeeb.Sharma-1@ou.edu}{}{University of Oklahoma}{}{Norman}{OK}{}{USA}

\begin{abstract}
The formation and evolution of disks early in the protostellar phase is an area of study
in which the ngVLA is poised to make significant breakthroughs. The high-sensitivity
and resolution at wavelengths of 3~mm and longer will enable forming disks to be examined
with unprecedented detail. The need to observe dust emission at wavelengths of 3~mm and longer
is motivated by the fact that dust emission at these wavelengths is more likely
to be optically thin, which is essential to understanding the structure of these
disks. We explore the feasibility of detecting and resolving protostellar
disks with a variety of radii, masses, and distances, out to distances as large as 1.5~kpc
using radiative transfer models and simulations with the proposed ngVLA configuration. We
also examine the potential for the ngVLA to enable studies of grain growth and
radial migration of dust particles early in the protostellar
phase with the broad multi-wavelength coverage. Studies of grain growth will require wavelength
coverage extending at least to $\sim$4~cm to characterize and quantify the location and intensity
of free-free emission, which is expected to be generated at $<$10 AU scales. 
\end{abstract}

\section{Introduction}

The formation of disks occurs as a natural consequence of angular momentum
conservation during the star formation process. As such, proto-planetary disks 
are found nearly ubiquitously toward pre-main sequence stars, with
higher fractions of disks found toward members of clusters/associations having 
younger collective ages \citep{hernandez2008}. The origins of proto-planetary
disks can be traced to the disks that form during the early phase of the 
star formation process, around Class 0 protostars. Class 0 protostars
are characterized by a dense envelope of infalling material that feeds the
protostellar disk. As the envelope dissipates due to continued accretion and
erosion of the envelope by protostellar outflows \citep[e.g.,][]{frank2014},
it becomes a Class I protostar. The Class 0 phase is expected to last $\sim$150~kyr,
while the combined Class 0 and Class I phases are expected to last $\sim$0.5~Myr \citep{dunham2014};
these ages estimates are based on the assumption that pre-main sequence stars
hosting proto-planetary disks (objects in the Class II phase) have an average age of $\sim$2~Myr. 
These young, embedded disks are often collectively referred to as protostellar disks.

Protostellar disks are important for the formation and mass
assembly of the star, as well as the planet formation process. Most of the 
mass that is accreted onto a star must pass through the disk, and 
these protostellar disks are the initial conditions for disk evolution
into a proto-planetary disk. The radius and mass of the forming
disk can be regulated by the angular momentum of the infalling material, and
magnetic fields can remove angular momentum efficiently in the absence of dissipative
processes \citep[e.g.,][]{li2014}. As such, the structure of these 
protostellar disks is expected to be connected to their formation conditions, and
clear detections of rotationally supported disks have been found toward 
Class 0 and I protostars \citep{tobin2012,harsono2014}. The disks may also increase in 
radius later via viscous evolution and the outward transport of angular 
momentum \citep[e.g.,][]{manara2016}. Furthermore, if gravitational 
instability in disks is a viable mechanism for angular momentum 
transport, the formation of multiple star systems, and possibly 
giant planet formation, the disks in the protostellar phase are
those most likely to have the requisite conditions \citep{kratter2010,tobin2016b}. 
Finally, the growth of solid material can
begin during the protostellar phase and this may catalyze 
the later formation of planets via the core accretion process
if the protostellar disk is able to efficiently grow dust grains to
pebble/rock sizes in order to promote planetesimal growth.

Typical proto-planetary disks have masses of $\sim$0.005~M$_{\odot}$, 
radii of $\sim$50~AU, and surface density profiles $\propto$ R$^{-1}$, 
meaning that most of the disk mass is at large radii \citep{andrews2013,ansdell2016}. 
Proto-planetary disks tend to be significantly
lower in mass than the protostellar disks, which can have 
masses $>$0.1~M$_{\odot}$ \citep{tobin2015,jorgensen2009,tychoniec2018}. The
typical masses, radii, and/or surface density profiles for protostellar
disks are still poorly constrained, and the ngVLA has significant potential to unlock some of these characteristics. 

\section{Resolving and Characterizing Youngest Disks}

\begin{figure}
\begin{center}
\includegraphics[scale=0.6]{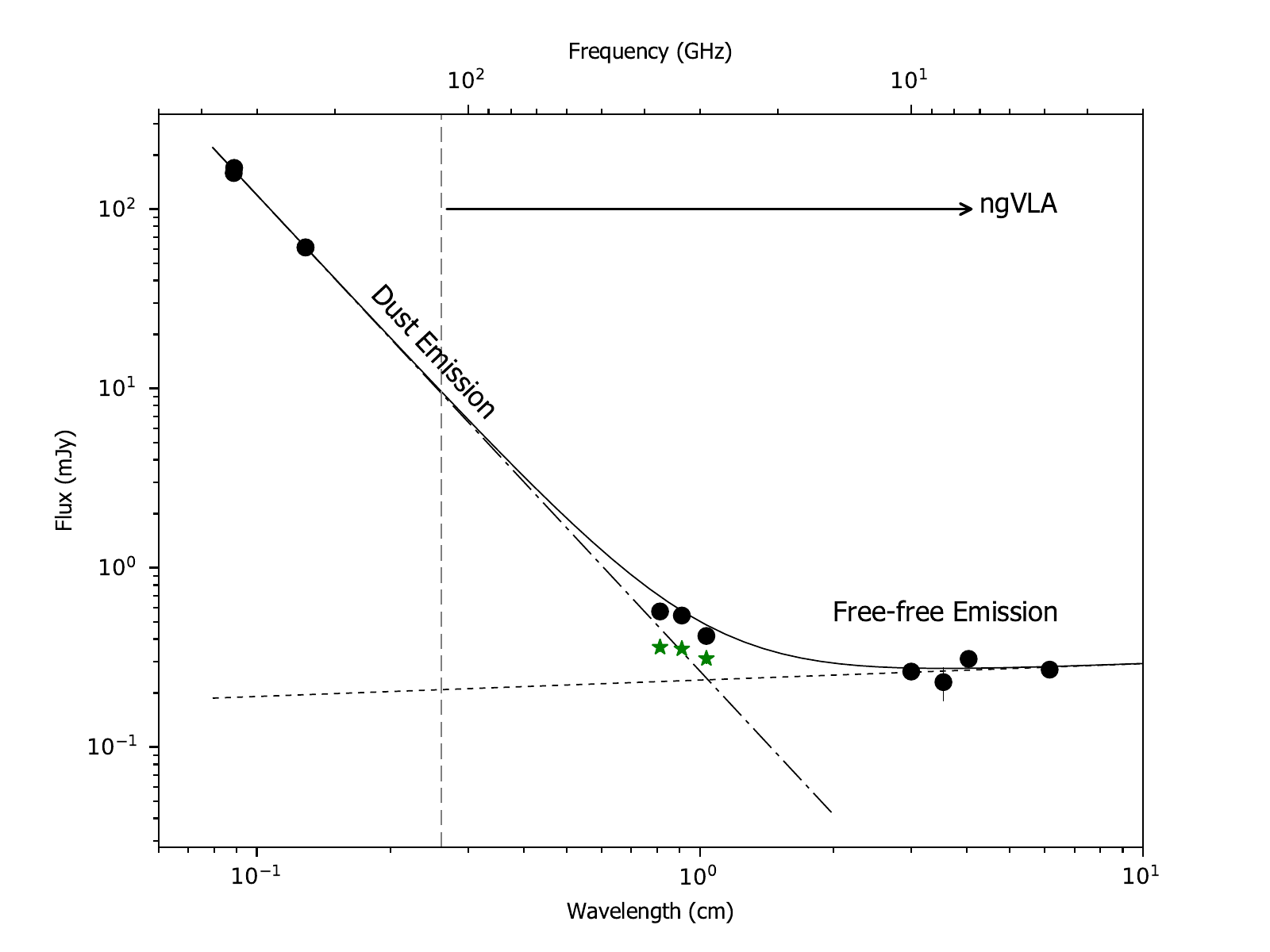}
\end{center}
\caption{Spectrum from millimeter to centimeter wavelengths of a typical 
protostar in Orion. The emission from dust dominates at short wavelengths
and free-free dominates at long wavelengths. The ngVLA will fully probe the transition
from dust emission to free-free. The 8~mm to 1~cm bands have some free-free emission that is
typically removed in current studies using power-law fits to the long wavelength 
emission. the ngVLA will enable improvements by spatially resolving the free-free emitting
region(s). The points at 0.87~mm and 1.3~mm are from ALMA, the points at all longer wavelengths
are from the VLA; the green star points are the flux densities between 8~mm and 1~cm with the
extrapolated free-free contribution subtracted. Error bars are plotted with each point, but in most cases the statistical uncertainties are smaller than the points plotted.
}
\label{spectrum}
\end{figure}

Observational studies of young disks have only just begun to reach samples 
larger than 10-20 systems \citep{tobin2015,seguracox2016}, 
and young protostar systems (Class 0 and Class I phases) 
are inherently more rare than pre-main sequence stars hosting
proto-planetary disks. This is because the lifetime of the
 protostar phase is expected to be at least $\sim$4$\times$
shorter than the lifetime of a PMS star with a disk \citep{dunham2014}. 
Specifically, the \textit{Spitzer} \textit{cores2disks} survey along 
with the entire \textit{Spitzer} Gould Belt survey contain 
only $\sim$200\footnote{This number corresponds 
to sources with the most firm characterization based on bolometric temperature.}
Class 0 and I protostars. Thus, to observe 
large numbers of protostellar disks, more populous star forming
regions need to be observed and these are only present at distances $\ge$ 400~pc. For example,
Orion alone contains $\sim$315 Class 0 and I protostars. The need to examine protostellar
disks at 100s of pc distances makes their characterization difficult to achieve with
the same mass sensitivity and spatial resolution as the 
more nearby proto-planetary disk systems. Moreover, high
mass sensitivity is particularly difficult at long wavelengths where the dust emission is intrinsically fainter (F$_{\lambda}$~$\propto$~$\lambda^{-(2+\beta)}$, where $\beta$ is the 
dust opacity spectral index). Figure \ref{spectrum} shows an example radio spectrum of a protostar
from the submillimeter to centimeter wavelengths which illustrates 
the expected spectral properties of protostar systems as a function of wavelength. Protostars 
often exhibit free-free emission that dominates at wavelengths $>$1~cm, but can can
contribute to the overall flux density at 7.5~mm. Thus, multi-wavelength observations
are important to measure and remove the contribution of free-free emission to the
flux density at wavelengths where we aim to trace dust emission. 
We will discuss this aspect further in section 4.
For the sake of discussion, throughout this chapter we will
use the dust opacity normalization of
0.899 cm$^2$~g$^{-1}$ (dust only) at 1.3~mm from \citep{ossenkopf1994}, assuming $\beta$=1 for extrapolation to longer wavelengths (0.155 cm$^2$~g$^{-1}$ at 7.5~mm).

\begin{figure}
\begin{center}
\includegraphics[scale=0.4]{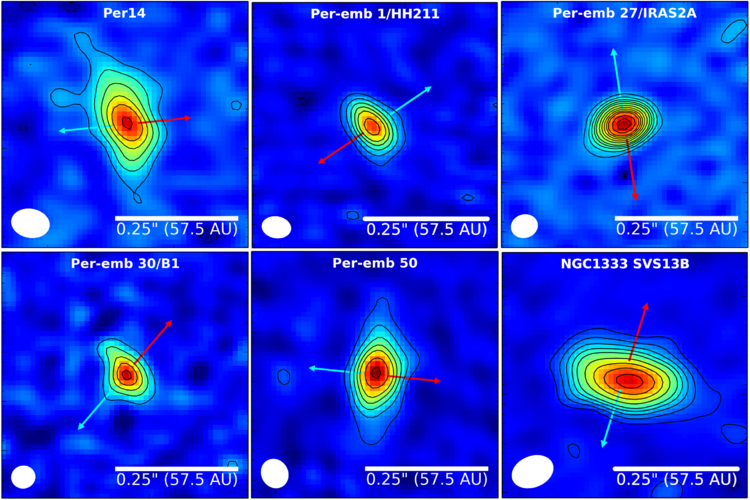}
\end{center}
\caption{Images of some well-resolved disks from the VANDAM:~Perseus survey
at 8~mm, from \citet{seguracox2016}.
The contours in each panel are [-6, -3, 3, 6, 9, 12, 15, 20, 25, 
30, 35, 40, 50, 60, 70, 80, 90, 100, 150] $\times$ $\sigma$, where
$\sigma$ = 11 \microjy\ at 8 mm.}
\label{perseus-disks}
\end{figure}

The advent of the Karl G. Jansky Very Large Array (VLA) with its enhanced continuum sensitivity has
enabled surveys of protostellar disks at wavelengths between 6.7~mm and 1~cm.
Dust emission is more optically thin than
at shorter wavelengths and thus the emission better traces column density due to the
power-law decrease of dust opacity with increasing wavelength \citep{hildebrand1983}.
Two VLA Nascent Disk and Multiplicity (VANDAM) surveys have been carried out 
with the VLA toward the Perseus and Orion star forming regions, observing a
total of over 200 protostar systems in each region. 
The VANDAM surveys took $\sim$600 hours of time on the 
VLA. These surveys were conducted in Ka-band at a central wavelength of 9~mm, while observing at the 
highest possible angular resolutions, translating to spatial resolutions of $\sim$15~AU and
$\sim$30~AU in Perseus and Orion, respectively. Example images of disks observed 
with the VLA are shown in Figure \ref{perseus-disks}. Also note that the VANDAM:~Perseus 
survey observed the entire sample at 4.1~cm and 6.4~cm, but we focus on the dust emission 
at 5.9~mm to 10~mm in this chapter. This corresponds to the proposed ngVLA band that is most 
similar to the wavelengths used in VANDAM and is centered at $\sim$7.5~mm.

\begin{figure}
\begin{center}
\includegraphics[scale=0.5]{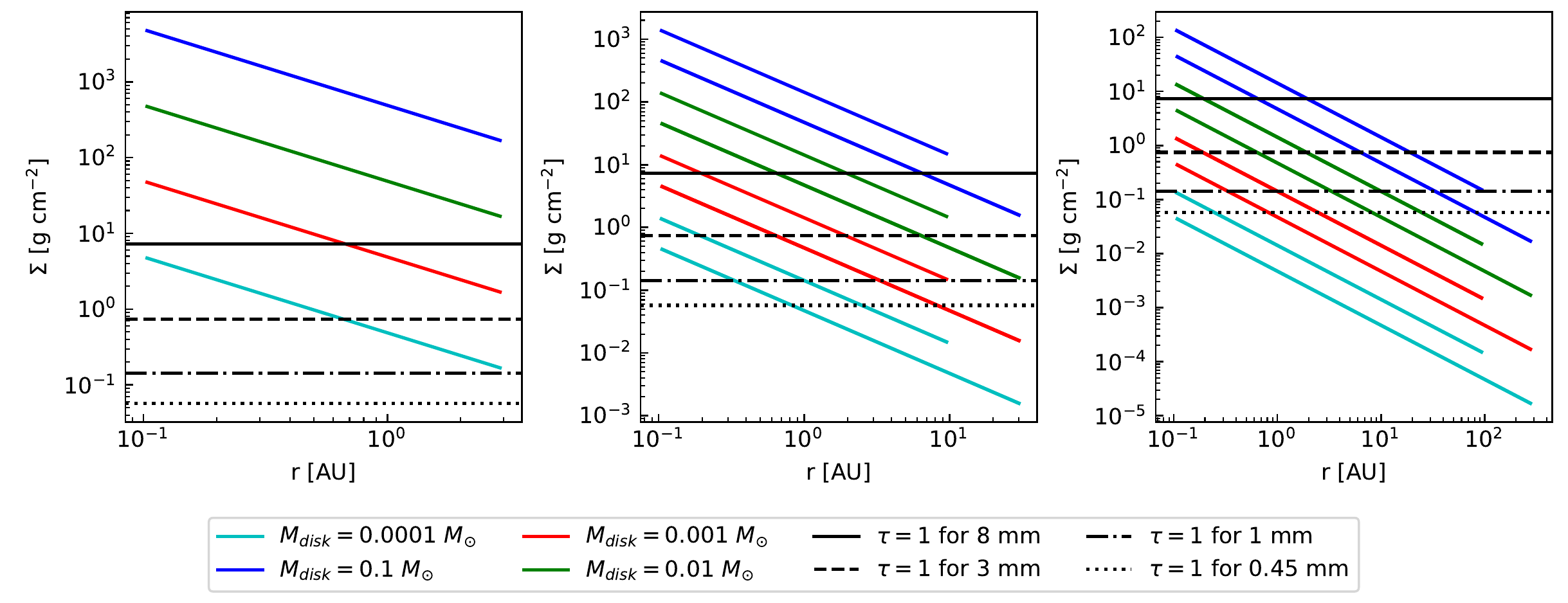}
\end{center}
\vspace{-5mm}
\caption{Surface density profiles of modeled disks versus radius. Note that range of radius
plotted increases from left to right; the 3~AU disks are plotted in the left panel,
the 10~AU and 30~AU disks are plotted in the middle panel, and the 100~AU and 300~AU disks
are plotted in the right panel. Horizontal lines are plotted
that denote the surface density corresponding to $\tau$~=~1 at 7.5~mm (solid), 3~mm (dashed), 1~mm (dash-dot), and 0.45~mm (dotted). Observations at 7.5~mm will not be immune to dust opacity effects, but
they can probe surface densities 10$\times$ larger without becoming optically thick. The dust mass
opacities calculated are 0.15, 1.35, 7.03, 17.43 cm$^2$~g$^{-1}$ at wavelengths of 7.5, 3, 1, 
and 0.45~mm, respectively.}
\label{opacity}
\end{figure}

The VANDAM:~Perseus survey detected candidate disks, meaning that extended dust continuum 
emission was resolved, toward only 12/42 Class 0 protostars and even fewer toward Class I 
protostars 5/37 (Segura-Cox et al. 2018 submitted). We show examples of some of the 
well-resolved disks detected
in the VANDAM:~Perseus survey in Figure \ref{perseus-disks} \citep{seguracox2016}. The disks
in Perseus were further characterized by \citet{tychoniec2018}, focusing on the integrated
dust emission at 9~mm toward all sources whether or not they were resolved. \citet{tychoniec2018}
found that Class 0 protostars tend to have more mass within radii $<$100~AU relative to the Class
I protostars. Furthermore, both Class 0 and Class I protostars had higher masses
than the more-evolved proto-planetary disks (Class II sources). 
The low percentage of well-resolved disks toward Class 0 and I protostars may be due to both
the limited spatial resolution and sensitivity of the VLA. The sensitivity may be especially 
limiting for Class I protostars as, due to their disk evolution, they 
have less overall mass (thus lower flux densities). The ability to observe at 
wavelengths of $\sim$9~mm is extremely important for examining protostellar 
disks, especially around Class 0 targets. If they are compact (R$<$ 50~AU) and 
massive ($>$0.01 M$_{\odot}$), they are likely optically thick at shorter wavelengths, but 
remain optically thin at most radii at 9~mm. We show radial surface density profiles for
model protostellar disks and their corresponding radii where $\tau$~=~1 in Figure 
\ref{opacity}, thereby underscoring that long wavelength observations
of dust emission are essential to resolve and quantify the majority of the mass within
protostellar disks. The opacity of disks at different wavelengths depending on 
mass and radius will be discussed in more detail in the following section.

Despite the advances enabled by the VLA surveys, they are not without limitations. Due to the increased distance to 
Orion, the VANDAM: Orion survey had 2.2$\times$ less mass sensitivity than the Perseus survey, despite
spending 2$\times$ more time on source. The VANDAM: Orion survey used 240 hours in the VLA 
A-configuration to observe 100 Class 0 protostars, accounting for $\sim$50\% of the
available array time at high-frequencies. Thus, high-sensitivity at 
$\sim$8~mm wavelengths toward large protostar samples
at a distance of $\ge$400~pc requires a significant utilization of
available observing time for the current VLA; the time required quickly becomes unrealistic 
for protostars at much greater distances. 

The ngVLA is needed to further advance studies of disks because 
ALMA will also not be as sensitive to long wavelength dust emission as the ngVLA. Using
ALMA at 3~mm, 1 hour of integration will achieve a sensitivity of 15~\microjy. Assuming
a distance of 230~pc, this translates to a 1$\sigma$ mass sensitivity of 
1.6$\times$10$^{-4}$~M$_{\odot}$. Therefore, ALMA Band 1 ($\sim$7.5~mm) will
only be about as sensitive as the current VLA. The ngVLA will reach 1$\sigma$
mass sensitivities of 5.0$\times$10$^{-5}$~M$_{\odot}$ and 8.2$\times$10$^{-6}$~M$_{\odot}$ at wavelengths of 7.5~mm and 3~mm,
respectively. Thus, the ngVLA will be vastly superior to ALMA
at these long wavelengths for the characterization of dust
emission from protostellar disks.
However, these mass estimates assume that all emission 
is within one beam. We will conduct a more
realistic assessment of the ngVLA's ability to detect resolved 
disks in the following section.

\section{Detectability of Protostellar Disks with the ngVLA}

The VANDAM:~Perseus survey was able to detect disks with total 
masses $>$ 0.025 M$_{\odot}$ (gas+dust, assuming a gas-to-dust ratio of 100:1 and
a 5$\sigma$ detection limit) 
and resolve disks with observed radii $>$10 AU at a wavelength of 8~mm (Segura-Cox et al. 2018).
However, most
of the resolved disks have masses $>$ 0.1~M$_{\odot}$. The 
sensitivity of the ngVLA will enable a $\sim$10$\times$ leap
in both resolution and sensitivity. In order to quantitatively characterize the ability
of the ngVLA to detect and resolve disks toward both nearby and distant star forming regions,
we have computed a small suite of radiative transfer models at wavelengths of 7.5~mm and 3~mm, varying 
luminosity (1.0, 10.0, 100.0 L$_{\odot}$), disk radius (3.0, 10.0, 30.0, 100.0, 300.0 AU), and disk
mass (10$^{-4}$, 10$^{-3}$, 10$^{-2}$, 10$^{-1}$ M$_{\odot}$).
The disks are assumed to have a radial surface density
profile proportional to R$^{-1}$, scale height (H/R) of 0.1 at 1~AU, and 
flaring H~$\propto$~R$^{1.15}$; the disk density profiles are truncated at the 
specified radii and do not exponentially drop-off. All models have a surrounding envelope with
a mass of 0.1~\msun\ and a radius of 1500~AU. The dust opacities we use are derived
from \citet{woitke2016} and have a maximum particle size of 1~mm, yielding dust
mass opacities of 0.15~cm$^2$~g$^{-1}$
and 1.35~cm$^2$~g$^{-1}$ at 7.5~mm and 3~mm, respectively.
Radiative transfer was computed with the RADMC3D code \citep{dullemond2012}, and model images were generated 
at three distances (230, 400, and 1500 pc) and three viewing geometries (i = 25, 45, and 75\deg).
We simulated ngVLA observations of these models
with the CASA \textit{simobserve} task using the full ngVLA configuration,
assuming 1 hour on source and the estimated noise of 
0.26~\microjy\ from this same integration time (ngVLA Memo \#17). 

\begin{figure}
\begin{center}
\includegraphics[scale=0.6]{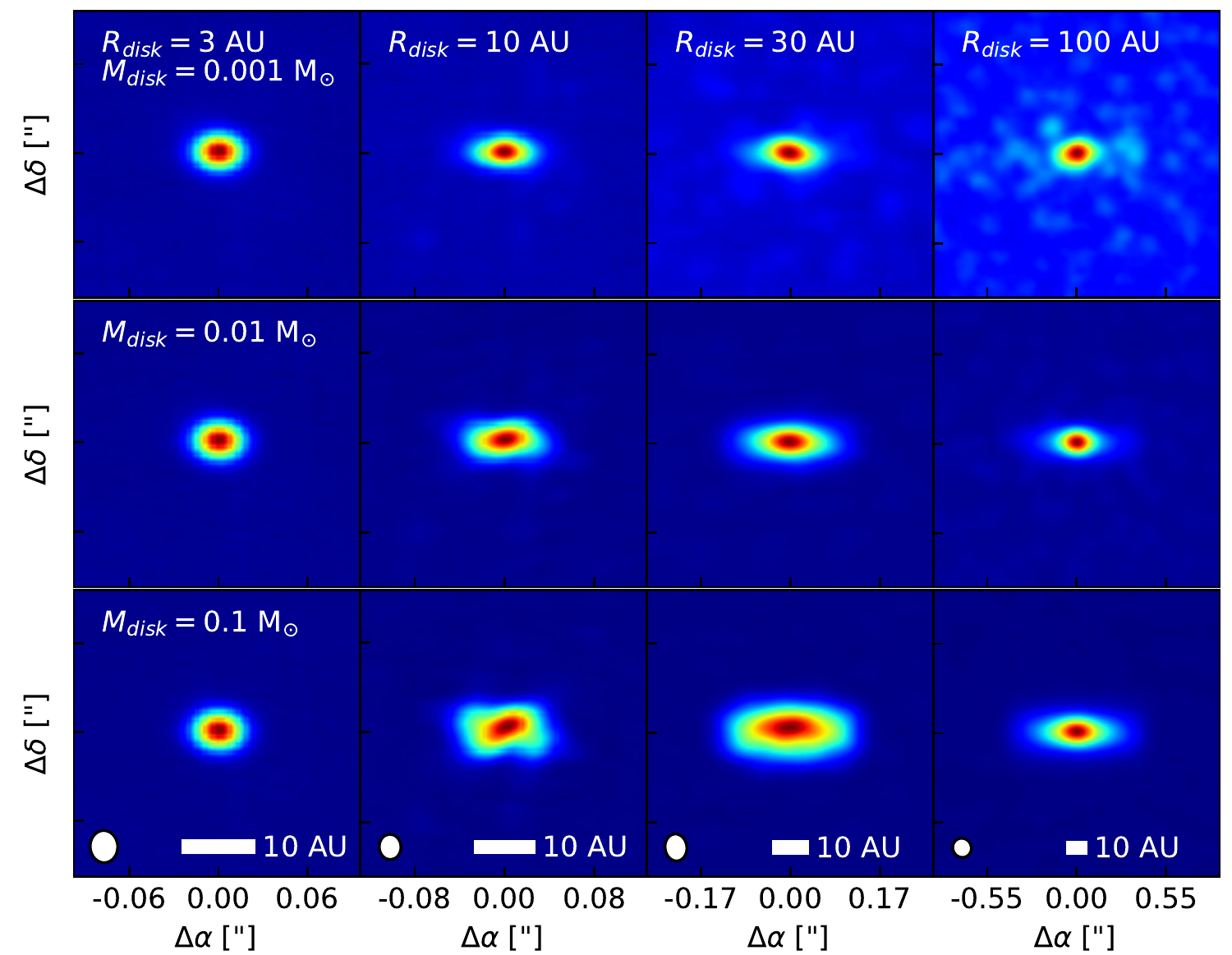}
\includegraphics[scale=0.6]{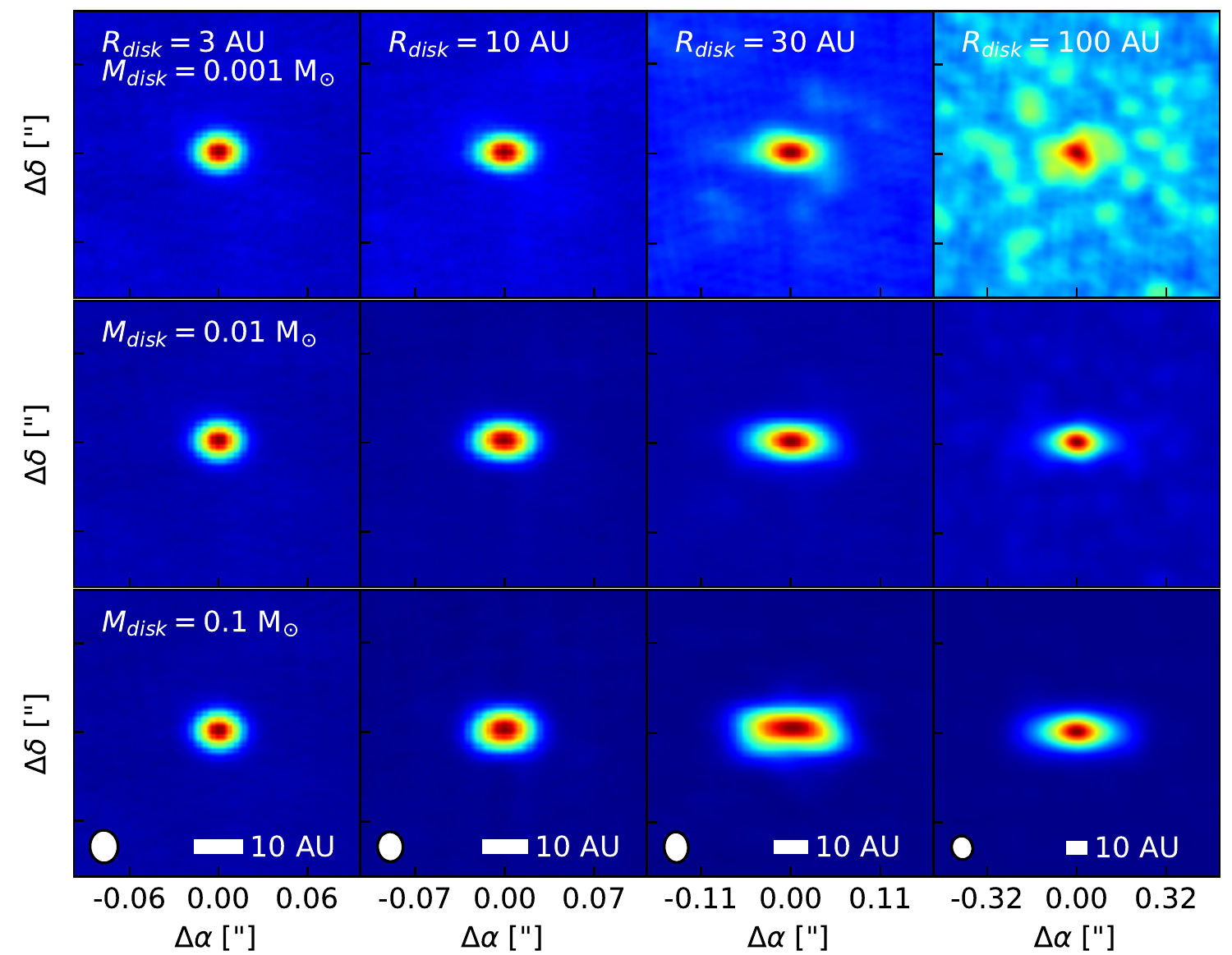}
\includegraphics[scale=0.6]{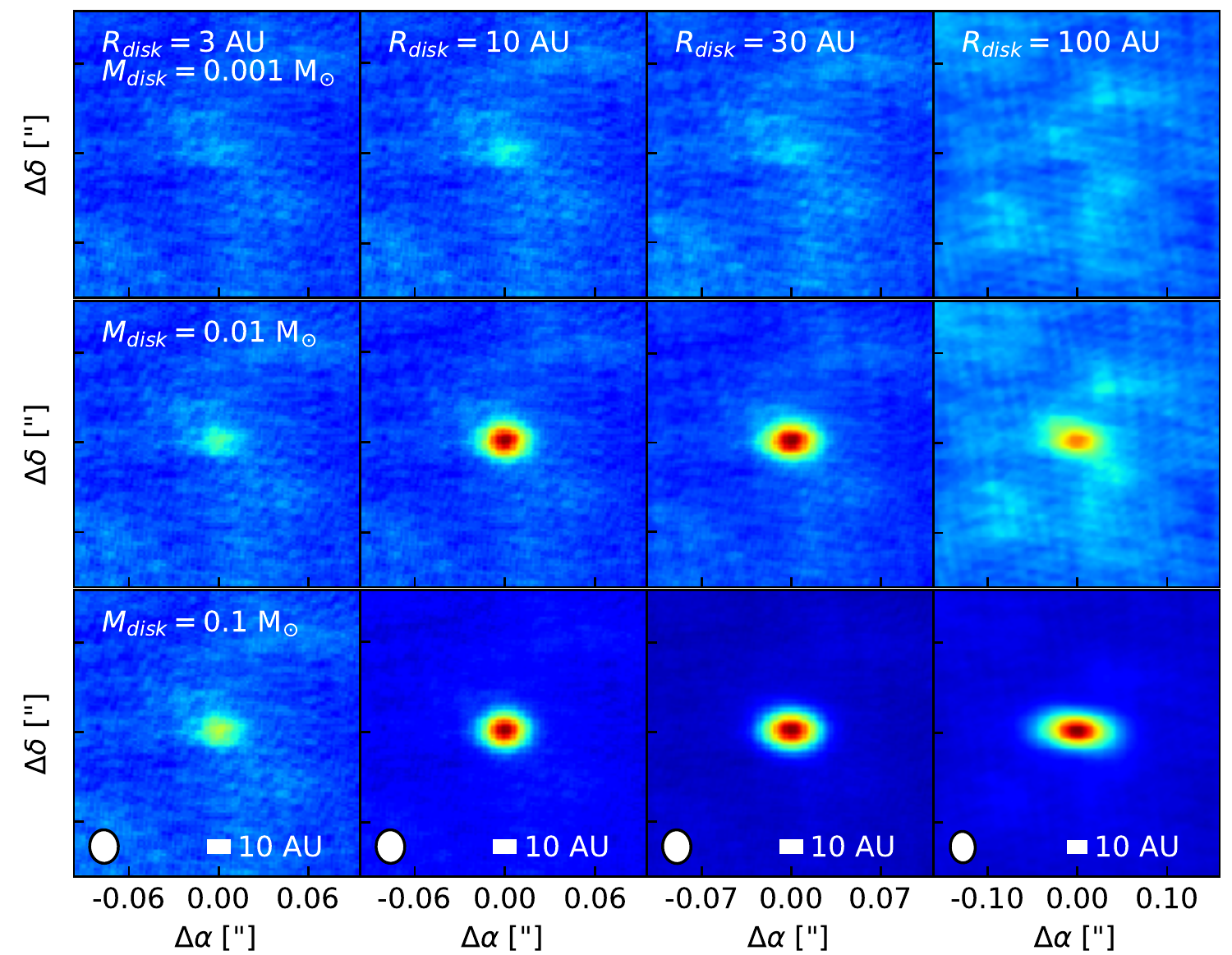}
\end{center}
\vspace{-5mm}
\caption{Model images of protostellar disks at 7.5~mm
computed for a three distances 230~pc (top), 300~pc (middle),
and 1500~pc (bottom). For each distance we have generated synthetic observations for a variety of radii and disk masses
and run through a simulated observation of 1~hr 
with the full ngVLA configuration. UV-tapering has been
applied to the disks with R$_{disk}$~$\geq$~30~AU at the level
of R$_{disk}$/5.0 translated to angular units.}
\label{disk-models}
\end{figure}

A subset of these models are shown in Figure \ref{disk-models} at a wavelength of 7.5~mm, 
distances of 230~pc, 400~pc, and 1.5~kpc, and at an inclination of 75\deg. At 
distances of 230~pc and 400~pc, disks 
with masses of $>$0.001 M$_{\odot}$ toward protostars with luminosities
of 10~L$_{\odot}$ can be detected and resolved. We note, however, that at a mass of 0.001~\msun,
the 300 AU radius disk is not as well detected due to the mass being spread over a larger 
disk size. We also find that disks with radii of 3 and 10~AU can also be detected for a disk
mass as small as 0.0001 M$_{\odot}$.

At a distance of 1.5~kpc and the same luminosity, disks with masses of 0.1~\msun\
can be detected and resolved at all radii, and at a mass of 0.01~\msun\ the 
disks with R$_{disk}$~$\leq$~100~AU can still be detected. Disks with masses much below 
0.01~$\msun$ are not well-detected in a 1 hour observation. While optimal imaging 
parameters will depend on the radii and mass of each
disk, the visibility data themselves can be utilized to fit the disk radii with
greater accuracy from the images alone, given the possible limitations of maximum recoverable
scale and surface brightness. The ngVLA will enable
the physical structure of protostellar disks with radii as small as $\sim$3~AU to be
characterized in the nearby star forming regions. The ngVLA is the only facility that will 
be able to do this given the necessity of very high angular resolution and
long-wavelengths; this will be a major advance
in the capability of examining small-scale structures in protostellar systems.

Furthermore, observations at 3~mm (not shown for brevity) enable the detection
of all the disks detected at 7.5~mm (Figure \ref{disk-models}), but 
lower mass disks will be able to be detected as well given the factor 
of $\sim$6 - 16$\times$ increase in flux density going from 7.5~mm to 3~mm
(depending on dust opacity spectral index). The 3~mm
band will be especially effective for the detection of disks in star forming
regions out to 1.5~kpc, as disk masses of 0.001~\msun\ can be detected
more robustly. Thus, both the 7.5~mm and 3~mm ngVLA bands will be essential to 
further characterizing disks in the protostellar phase, and the full ngVLA will enable disks to be detected and resolved
out to star forming regions at distances of at least 1.5~kpc. This capability will greatly
expand the number of disks that can be observed compared with the current VLA by
enabling significantly smaller radii and masses to be detected and resolved. By including
all star forming regions out to $\sim$1.5~kpc \citep[e.g., Cygnus-X;][]{kryukova2014}, more than 1000 additional protostars and their
disks will be accessible to the ngVLA. Moreover, the 100s of protostars star-forming regions 
at distances of 400~pc or less \citep{dunham2014} will be able to have 
their disks resolved, where only unresolved
observations were possible with the current VLA.

Optical depth is another significant advantage of the ngVLA over other millimeter 
interferometers like ALMA that primarily operate at wavelengths shorter than 3~mm. 
The surface density profiles for the modeled disks are shown in Figure \ref{opacity} 
with horizontal lines marking
the surface density corresponding to $\tau$~=~1 for wavelengths between 7.5~mm and 0.45~mm. Due to the 
fixed grid of masses and varying radii, the surface density profiles are larger for smaller disk radii. However, it can be seen that 7.5~mm wavelengths can trace about 10$\times$ higher surface densities than at 1~mm before becoming optically thick. The opacity
can be significant even at 7.5~mm, particularly for disks with M$_{disk}$~=~0.1~M$_{\odot}$. 
For the 100 and 300~AU radius and M$_{disk}$~=~0.1~M$_{\odot}$ disks, only the inner few AU are 
optically thick at 7.5~mm, while the entire 100~AU disk is optically thick at 1~mm. The disks with radii
between 10 and 30 AU and M$_{disk}$~=~0.1~M$_{\odot}$ are optically thick out to $\sim$10~AU in both
cases, but lower mass disks in this range of radii are only optically thick in their inner few AU.
However, these disks are completely optically thick at 1~mm and 0.45~mm until the disk masses are
$\sim$0.0001~M$_{\odot}$. Note that $\tau$ in Figure \ref{opacity} is calculated for an inclination of 0\deg\ (face-on), and
intermediate to edge-on inclinations will have substantially more opacity for a given surface density profile.
Therefore, we can conclude that 7.5~mm wavelengths are not necessarily
immune to issues of opacity, but they are significantly less affected than 1~mm and 0.45~mm. This will enable ngVLA
to significantly improve on the characterization of protostellar disks, while the ability of ALMA
to probe their structure is more limited.

The ability to detect these low-mass, small radius disks is also extremely
important for the characterization of multiple star systems from the emission 
of their circumstellar disks. This capability is
described in more detail in the Chapter `New Frontiers in Protostellar Multiplicity with the ngVLA.' Furthermore, while we do not specifically address the likelihood of 
substructure in protostellar disks, growing numbers of proto-planetary disks
exhibit a variety of substructure when observed with sufficiently high resolution
in the dust continuum \citep[][Andrews et al. 2018 in prep.]{brogan2015, andrews2016,sheehan2018}.
Therefore, it is entirely possible that protostellar disks themselves will also not have
smooth structures. The higher-resolution at longer wavelengths of the ngVLA
will be crucial for characterizing whether or not protostellar disks exhibit significant
substructure. This is because, as discussed previously, the more massive protostellar
disks are largely opaque at shorter wavelengths and the longer wavelengths with lower
opacity are more ideal to reveal substructure if present.

\section{Characterizing Early Grain Growth}

The ability to detect and resolve disks with a variety of radii in the nearby star forming
regions makes it possible to characterize the spectral index of dust emission both in 
the limit of integrated flux densities and spatially-resolved emission throughout the disks. With a shortest wavelength of 3~mm, the ngVLA can characterize 
the spectral index of dust emission at least between 3~mm and 1~cm. 
The sensitivity of the $>$1~cm 
wavelength ngVLA bands will also have the capability of detecting and resolving
dust emission. This capability is incredibly important
because the rate of growth for solids in disks can be examined as a function of evolution
through the protostellar phase. However, at wavelengths $>$1~cm dust emission becomes
more difficult to detect due to the power-law decrease in flux density of dust emission
at longer wavelengths, thus the ngVLA is crucial to enable the detection of dust emission
at wavelengths $>$1~cm because the current VLA does not have the needed angular resolution 
or sensitivity.

The ability to measure the spectral index of dust emission 
over a wide range of wavelengths is important because it is 
connected to both the maximum dust grain sizes and the slope of
the assumed power-law dust size distribution; dust grains are most efficiently detected
when observed at a wavelength comparable to their size.  Through radiative transfer 
modeling, using the same techniques that generated the images in Figure \ref{disk-models},
the dust emission over a wide range of wavelengths can be used
to determine the maximum particle sizes. Such studies are currently 
possible toward a few select systems, but the limited sensitivity 
of the VLA makes detection of dust emission out to large disk
radii and low disk masses challenging. Furthermore, the low angular
resolution of the current VLA at long wavelengths means that characterizing 
dust emission a wavelengths of several cm must use integrated 
flux densities. Resolved studies of disk spectral indices 
have been done for a few of the most nearby disk-hosting PMS 
stars \citep{perez2012,perez2015}. Doing such modeling for protostars spanning the
full range of evolution will enable the rate of grain growth to be understood, and 
we will learn whether or not this is a linear/universal process, or if it varies strongly
between protostars. However, in order to accomplish this, greater 
sensitivity and resolution than are possible with the VLA are needed to sufficiently 
resolve the inner regions of protostellar disks and accurately separate 
free-free emission from dust emission.

Separating the free-free jet emission from dust emission is incredibly important, especially
for examining dust emission at $>$1~cm (Figure \ref{spectrum}). The resolution of 
the ngVLA will enable the dust emission to be resolved from the free-free 
emission out to wavelengths of several cm. However, the disk needs to be well-resolved
for this to be possible. Compact free-free emission is expected to originate from the
inner $\sim$10~AU \citep{anglada1998} and the ngVLA will have a spatial
resolution of $\sim$11.5~AU in the 3.75~cm (8 GHz) band for protostars at a distance of 230~pc. 
Thus, the emission from the two distinct processes 
can be separated spatially with multi-wavelength observations, 
and the region containing free-free emission can be 
isolated and removed from the analysis. This is not possible with the angular resolution
of the current VLA and will be an unique capability of the ngVLA. 
We also note that free-free emission
can also originate from a disk wind, but without the ngVLA we cannot fully characterize the 
nature of free-free emission to properly account for it. Finally,
free-free emission is can still contribute at wavelengths $<$1~cm (Figure \ref{spectrum}), and the higher angular
resolution at shorter wavelengths will help further characterize its nature.

ALMA can add to the characterization of dust growth with its equivalent angular
resolution at shorter wavelengths. While small disks and the inner radii
of large disks will be optically thick (Figure \ref{opacity}), ALMA observations should
be more sensitive to the outer regions of protostellar disks where 
the concentration of large grains is expected to be lower. This is because
grain growth is slower at larger radii, having lower densities, and the radial drift process
is expected to move large particles to smaller radii \citep{weidenschilling1977}. 
Therefore, in the regions where the dust emission is not optically thick, short wavelength
observations from ALMA will be helpful to characterize dust growth at all disk radii, 
especially if the dust emission is more compact at longer wavelengths \citep{perez2012,perez2015,seguracox2016}.

Furthermore, the ngVLA may not be able to probe the transition from 
the envelope to disk as well as ALMA, because
the envelope will have significantly lower column density and thereby low surface 
brightness. On top of that, if the envelopes have mostly small dust 
grains $<$ 10~$\micron$, the dust emission
drops much faster in wavelength than it does in the disk due to the smaller grains. Thus,
ALMA will enable the connection of the disk to the infalling/rotating envelope to 
be better understood through both the observations of shorter 
wavelength dust and molecular lines tracing 
envelope and disk kinematics. The ngVLA then will be the ideal
instrument for examining the
presence and structure of the disks around protostars to smaller radii. This will complement
ALMA's ability to examine the structure of proto-planetary disks
around more evolved young stars.

\section{Uniqueness of ngVLA Capabilities}

The ngVLA will have extraordinary capabilities to examine disks toward protostars throughout
the Class 0 and I phases. It is crucial to examine the disks at long wavelengths where the dust
emission is optically thin over a large range of radii to probe nearly the entire disk mass reservoir. Moreover, the ngVLA
will be able to fully resolve young disks with radii as small as $\sim$3~AU in the
nearby star forming regions. While ALMA can reach such spatial scales
at short wavelengths, 
the disks around protostars are likely to be optically thick at 
such small radii (Figure \ref{opacity}), making it impossible
for ALMA to examine the internal structure of their protostellar disks in many cases.

Due to the optical depth of disks (Figure \ref{opacity}), the radial distribution of dust particle sizes cannot
be explored with short wavelength data from ALMA. Longer wavelength observations and $\sim$10~mas
resolution are crucial to map the radial distribution of dust emission. While ALMA has 3~mm 
receivers and is expected to have $\sim$1 cm receivers (Band 1) 
in the future, the best angular 
resolution at 3~mm is $\sim$0.05\arcsec, and $\sim$0.14\arcsec\ at 1~cm. 
Thus, ALMA at 3~mm will only have the spatial resolution that VANDAM had with the VLA
toward Perseus at 8~mm, and ALMA will have lower angular resolution at 1~cm; ALMA will not
be able to improve upon the resolution of the VANDAM survey at wavelengths
where dust emission is less affected by opacity. Furthermore, ALMA will be 
significantly less sensitive than the ngVLA at these same 
wavelengths; see section 2. In summary, the ngVLA will be 
superior to ALMA in examining protostellar disks at wavelengths of 3~mm and longer.

Lastly, the ngVLA will be able to examine the vertical settling of large 
dust grains for edge-on disks. In nearby star forming regions, 
(e.g., Taurus, Oph at 140 pc), edge-on disks can be
observed to determine how settled the disk midplanes are. The disk midplanes 
can be examined from 3~mm to 1~cm to see if there is variation with wavelength 
in the vertical extent of the dust emission. 
If the vertical structure of the dust emission is approximately H/R $\sim$0.1, then at a radius
of 14~AU the vertical height of the dust at 7.5~mm can be resolved 
(assuming 0.01\arcsec\ resolution at 140~pc). This study also cannot be 
done with ALMA at shorter wavelengths because the
disk midplanes are opaque \citep{lee2016}. Through modeling, 
it can be determined if the dust emission is 
more concentrated in the disk midplanes going to longer wavelengths, consistent with expectations of dust settling \citep{dalessio2006} and low levels of 
turbulence in the disks (e.g., Flaherty et al. 2017).

\section{Summary}
The ngVLA will enable a new revolution in the study of protostellar disks with the ability
to probe the spatial structure of protostellar disks, where many of these disks cannot be
resolved with current instrumentation. The ngVLA will enable us to resolve dust emission
from disks that are as small as 3~AU in the nearby star forming regions 
and examine the full spatial extents of protostellar disks using emission from
optically thin dust. Together, this information will enable us to determine the 
spectral index and thereby the sizes of the emitting
dust grains to be characterized as a function of radius, and perhaps height for edge-on disks.
The ngVLA will enable the structure and early grain growth of protostellar disks to be better
understood, in addition to how quickly the conditions for planet formation are established in 
disks.



\bibliography{tobin_disks}  

\end{document}